\documentclass[twocolumn,secnumarabic,amssymb, nobibnotes, aps, prd,superscriptaddress]{revtex4-1}

\usepackage{graphicx}
\usepackage{multirow}
\usepackage{caption}
\usepackage{subcaption}
\usepackage{upgreek}

\begin{document}

\title{High-order harmonic generation using a high-repetition-rate turnkey laser}

\author{E. Lorek}%
\email{eleonora.lorek@fysik.lth.se}
\affiliation{Department of Physics, Lund University, Box 118, 221 00 Lund, Sweden}
\author{E. W. Larsen}
\affiliation{Department of Physics, Lund University, Box 118, 221 00 Lund, Sweden}
\author{C. M. Heyl}
\affiliation{Department of Physics, Lund University, Box 118, 221 00 Lund, Sweden}
\author{S. Carlstr\"{o}m}
\affiliation{Department of Physics, Lund University, Box 118, 221 00 Lund, Sweden}
\author{D. Pale\v{c}ek}
\affiliation{Department of Chemical Physics, Lund University, Box 124, 221 00 Lund, Sweden}
\author{D. Zigmantas}
\affiliation{Department of Chemical Physics, Lund University, Box 124, 221 00 Lund, Sweden}
\author{J. Mauritsson}
\email{johan.mauritsson@fysik.lth.se}
\affiliation{Department of Physics, Lund University, Box 118, 221 00 Lund, Sweden}
\date{September 8, 2014}%

\begin{abstract}
We generate high-order harmonics at high pulse repetition rates using a turnkey laser. High-order harmonics at 400 kHz are observed when argon is used as target gas. In neon we achieve generation of photons with energies exceeding 90\,eV ($\sim$13\,nm) at 20\,kHz. We measure a photon flux of 4.4$\cdot10^{10}$ photons per second per harmonic in argon at 100\,kHz. Many experiments employing high-order harmonics would benefit from higher repetition rates, and the user-friendly operation opens up for applications of coherent extreme ultra-violet pulses in new research areas.

\end{abstract}

\maketitle

\section{Introduction}
\label{Sec: Introduction}
Coherent extreme ultra-violet (XUV) pulses, providing excellent temporal and spatial resolution, can be produced by a process called high-order harmonic generation (HHG) \cite{FerrayJPB1988}. The harmonic spectrum typically consists of odd harmonic orders of the driving field frequency that form a plateau extending over a large energy range up to high photon energies \cite{PopmintchevScience2012}. By isolating a single harmonic order, high spatial resolution can be obtained~\cite{BartelsScience2002}. While an isolated harmonic order has a duration on the order of the driving field, the total harmonic spectrum corresponds to a train of pulses with hundreds of attoseconds in duration \cite{PaulScience2001}. By applying various temporal confinement schemes isolated attosecond pulses can also be obtained \cite{HentschelNature2001}. These attosecond pulses have been fruitfully used, {\it e.g}. for studies of the motion of electrons on their natural timescale~\cite{DrescherNature2002,CavalieriNature2007,MauritssonPRL2008,SansoneNature2010,SchulzeScience2010,MauritssonPRL2010,KrauszRMP2009}.

High-order harmonics are produced by focusing a laser into a gas. The process is usually described with a simplified semi-classical three-step model \cite{CorkumPRL1993, SchaferPRL1993}, where the electric field distorts the Coulomb potential of the atom allowing an electron to 1) tunnel out; 2) be accelerated by the electric field; and 3) recombine with its parent ion, leading to emission of coherent XUV light. Since this process is repeated every half-cycle of the laser field, interference between consecutive pulses results in odd high-order harmonics of the fundamental frequency and, in the time-domain, an attosecond pulse train (APT). By increasing the duration of the driving field the number of pulses in the APT is increased, leading to a spectral narrowing of the harmonic orders allowing for an improved spectral resolution while the pulse duration of the individual attosecond pulses may be unaffected, thereby maintaining the possibility for very high temporal resolution.

HHG requires laser intensities on the order of $10^{14}$\,W/cm$^2$ to distort the Coulomb potential and allowing the electron to tunnel out and get accelerated to high energies. Such fields are usually obtained by  energetic pulses with short time durations, produced by large high power laser systems. Due to the limitations in average power of such systems, high energy pulses are obtained at the cost of repetition rate. Traditionally, HHG has therefore been limited to low pulse repetition rates, not exceeding a few kHz. Many experiments using HHG would, however, benefit from higher repetition rates. Examples include time-resolved photoemission electron spectroscopy and microscopy, two techniques that are limited in the number of emitted photoelectrons per pulse by space charge effects and therefore need a higher repetition rate in order to provide better statistics. Other experiments where coherent XUV pulses at a high repetition rate are needed include coincidence measurements \cite{Ullrich03rpp, Singh2010PRL, SchumannPRL2007} and the generation of frequency combs in the XUV range for high-resolution spectroscopy \cite{GohleNature2005, KandulaPRL2010, JonesPRL2005}.

The fact that high power laser systems are used for HHG has not only limited the repetition rates but also the availability to relatively few laboratories with the right equipment and expertise to run the systems. Generating high-order harmonics with a more compact and user-friendly laser system than those traditionally used, could increase the availability. Examples of fields potentially benefiting from accessible HHG include, apart from attosecond science itself, femtochemistry, nanophysics \cite{MikkelsenRSI2009} and surface science \cite{HuthAPL2014}. 

\begin{figure*}[bth]\centering
 \includegraphics[width=0.7\linewidth]{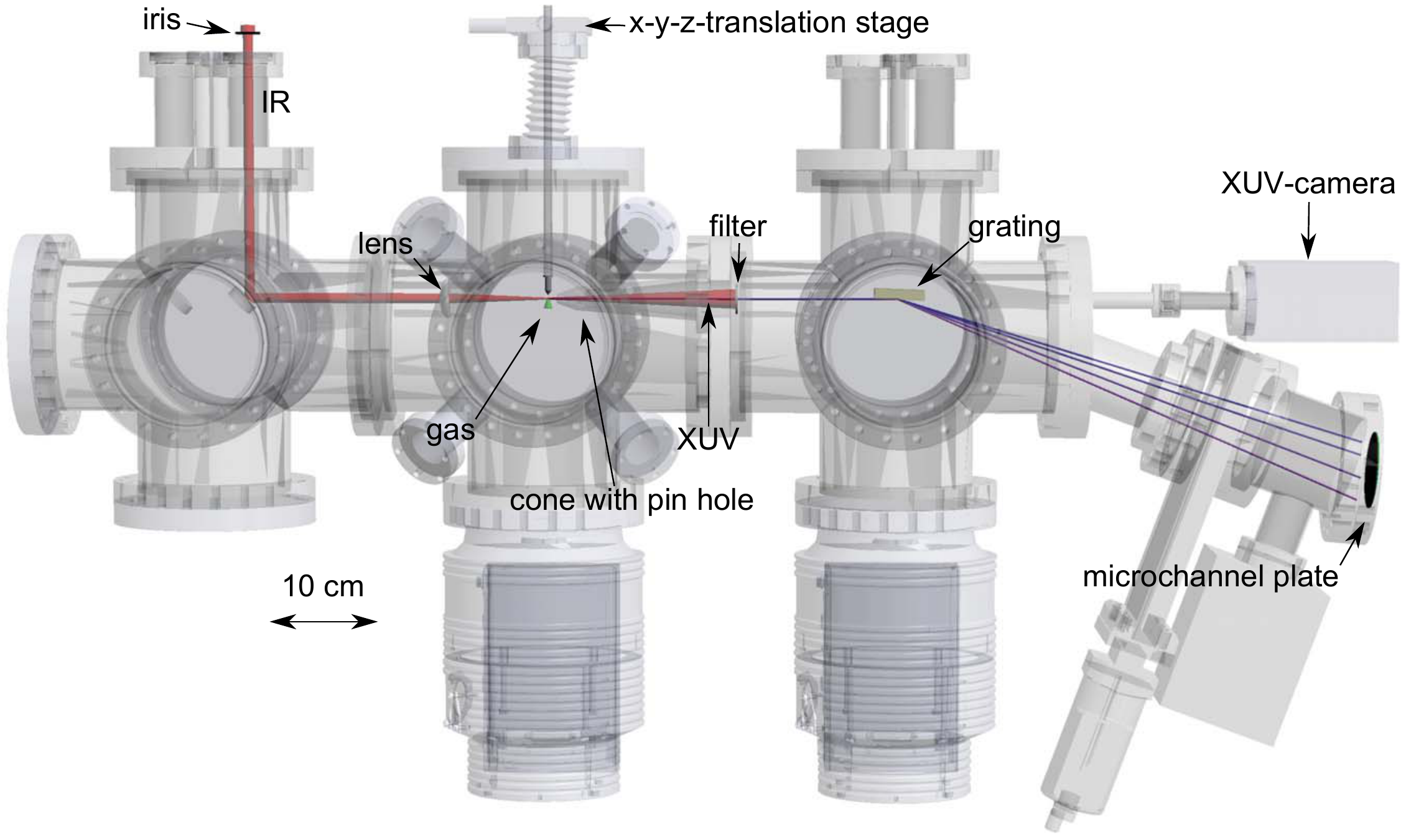}
  \caption{The experimental HHG setup including the beam path. The beamsizes are exaggerated. To allow room for a future interferometer enabling attosecond pump-probe experiments, the beam path is off-centered. A full 3D model of the HHG setup in pdf-format can be found as supplementary material. (Multimedia view)}
  \label{fig: Setup}
\end{figure*}

Both the issue of the traditionally limited repetition rates and the issue of limited accessibility of HHG light are addressed in this paper. Several recent studies report the generation of high-order harmonics at high-repetition rates (50\,kHz--25\,MHz) \cite{LindnerPRA2003, ChenOE2009, BoulletOL2009,  HadrichOE2011,  VernalekenOL2011, ChiangAPL2012, KrebsNP2013, FuchsRSI2013, HeylJPB2012, RothardtNJP2014, GohleNature2005, JonesPRL2005, YostNphys2009, KimNat2008, HuthAPL2014}. The general approach is to use a laser system with a high repetition rate, either an oscillator or an amplified system. In order to achieve the required intensity for HHG, despite the low pulse energy, three schemes have been applied. The first scheme \cite{LindnerPRA2003, ChenOE2009, BoulletOL2009,  HadrichOE2011,  VernalekenOL2011, ChiangAPL2012, KrebsNP2013, FuchsRSI2013}, investigated in detail by \cite{HeylJPB2012, RothardtNJP2014}, uses a tight-focusing geometry where the laser light is focused by a mirror or lens with short focal length onto the gas medium, effectively producing a high intensity at the focus. The second scheme \cite{GohleNature2005, JonesPRL2005, YostNphys2009} uses a high-finesse optical resonator that contains the medium for HHG. In this intra-cavity scheme, high-order harmonics are produced at the high repetition rate of the seed oscillator. This technique is, however, experimentally challenging. The third scheme exploits the local field enhancement induced by resonant plasmons within a metallic nanostructure to achieve the intensity needed for HHG \cite{KimNat2008}. If this is actually a viable route to generate harmonics has, however, been questioned \cite{SivisNature2012}.

In a recent work \cite{HuthAPL2014}, not only the traditionally limited repetition rates for HHG was addressed but also the accessibility of HHG light. HHG with energies between 13\,eV and 45\,eV from a turnkey laser at repetition rates of 0.2--25\,MHz is reported.

In this work we perform a more detailed investigation of HHG using a high-repetition-rate turnkey laser. Using this compact, stable and user-friendly laser in a tight focusing-scheme, we can achieve generation up to 92.7\,eV (13.4\,nm) at 20\,kHz in neon. An XUV photon flux of up to 4.4$\cdot10^{10}$ photons/s per harmonic order using argon and 100\,kHz is measured. We can generate harmonics 
at up to 400\,kHz repetition rate in argon.

In section \ref{Sec: Experimental setup} the experimental method and setup are presented. Section \ref{Sec: Results} is devoted to a presentation of our results followed by a conclusion in Section \ref{Sec: Conclusion}.

\section{Experimental method and setup}
\label{Sec: Experimental setup}
The main challenge with HHG at high repetition rates is the limited energy per pulse, setting constraints for reaching the intensities needed while keeping macroscopic generation conditions optimized. The three schemes mentioned above address this limitation in different ways, using power or intensity enhancement approaches in combination or solely with a tight-focusing geometry. In this work, we employ the straightforward tight-focusing approach in a single-pass configuration.  

The tight-focus geometry leads to a small interaction volume, demanding a high gas density for efficient HHG \cite{HeylJPB2012}. Confining a high density gas target to a small interaction volume while limiting the backing pressure in the vacuum chamber in order to avoid re-absorption of the generated XUV radiation, is technically challenging \cite{RothardtNJP2014}. The tight focusing also leads to a very divergent harmonic beam, which makes the manipulation and transport of the beam for further experiments more difficult. It is therefore desirable to focus the fundamental laser beam as loosely as possible, while still reaching sufficient generation intensities. The necessary laser intensity, $I_L$, can be estimated from the cut-off law \cite{KrausePRL1992}, which defines the highest harmonic order $q$ of the plateau harmonics that typically can be observed: 
\begin{equation}
\hbar q \omega = I_p + 3.17 U_p.
\end{equation}
Here, $\omega$ is the laser angular frequency, $I_p$ the ionization potential of the atoms used for HHG and $U_p \sim I_L \lambda^2$ the ponderomotive energy with $\lambda$ denoting the laser wavelength. The cut-off law predicts that the ponderomotive energy, and thereby the cut-off energy, can be increased by increasing either $I_L$ or $\lambda$. However, it has been shown that the efficiency scales very unfavorably with increasing wavelength~\cite{TatePRL2007}. Nevertheless, increasing the driving wavelength from the typically used 800\,nm to 1030\,nm allows us to relax the intensity requirement while still reaching high cut-off energies and good conversion efficiencies at high repetition rates.

The laser used to drive HHG is a commercially available ("PHAROS", Light Conversion), compact Yb:KGW  laser ($640\times\,360\times\,212$\,mm), which is both user-friendly and stable in its operation (pulse energy, pointing and pulse duration). The wavelength is 1030\,nm and the repetition rate is tunable between 1\,kHz and 600\,kHz and, in addition, a number of pulses can be removed using a pulse picker. With changed repetition rate, the pulse energy can be varied between 0.5\,mJ and 10\,$\upmu$J. Note that the average power is not constant over the tunability range. The pulse duration is 170\,fs.

The experimental setup, including the beam path and the vacuum chambers, is presented in Fig.\,\ref{fig: Setup} (Multimedia view). The laser light is sent into the chamber and focused by an achromatic lens with 100\,mm focal length. The target gas (argon, neon or air) is supplied by an open-ended, movable gas nozzle with 90\,$\upmu$m inner diameter. The generated XUV light then propagates, together with the IR light, through a small pinhole placed close after the generation location. The pinhole separates the generation part of the chamber from the detection part, each pumped by a turbo pump with a capacity of 500\,l/s (Oerlikon Leybold vacuum, MAG W600). The small pinhole allows for a very steep pressure gradient with up to $10^{-2}$\,mbar in the generation part of the chamber while maintaining $10^{-7}$\,mbar in the detection part. After the generation the IR light is removed by a 200\,nm thin aluminum filter.

The generated harmonic emission is detected using a flat-field grazing-incidence XUV spectrometer based on a blazed, varied-line-space XUV grating (Hitachi, Grating 001-0639) with 600\,lines/mm, which is placed on a rotation stage and a linear translation stage. The grating diffracts and focuses the harmonics along the dispersive plane and reflects them in the perpendicular direction onto a 78\,mm diameter microchannel plate (MCP) (Photonis). This allows us to study both the spectral contents of the emission and the divergence of the individual harmonics. The MCP is finally imaged by a CCD camera (Allied Vision Technologies, Pike F-505B). By removing the grating from the beam path with the linear stage the photon flux can be measured in the forward direction using an X-ray camera (Andor, iKon-L DO936N-M0W-BN). The spectrometer was calibrated using the grating equation with the known geometry of the setup and the grating constant as inputs. The calibration was verified by two atomic lines in argon and two atomic lines in neon. The argon lines used are 104.82\,nm \cite{VelchevJPB1999} and 106.67\,nm \cite{MinnhagenJOSA1973} corresponding to the transitions $3p^5(^2P_{1/2})4s\rightarrow3p^6(^1S_o)$ and $3p^5(^2P_{3/2})4s\rightarrow3p^6(^1S_o)$, while the two neon lines are 73.590\,nm and 74.372\,nm corresponding to transitions $2s^22p^5(^2P^o_{1/2})3s\rightarrow2s^22p^6$ and $2s^22p^5(^2P^o_{3/2})3s\rightarrow2s^22p^6$ \cite{SalomanJPPCRD2004}.

\section{Results}
\label{Sec: Results}

Using the experimental setup described in Section \ref{Sec: Experimental setup}, high-order harmonics were successfully generated in both argon and neon, but conveniently also in air (mainly nitrogen). The use of air at a backing pressure of 1\,atm is a very convenient experimental trick as it does not require any gas bottles or connections.

\begin{figure}
\includegraphics[width=9cm]{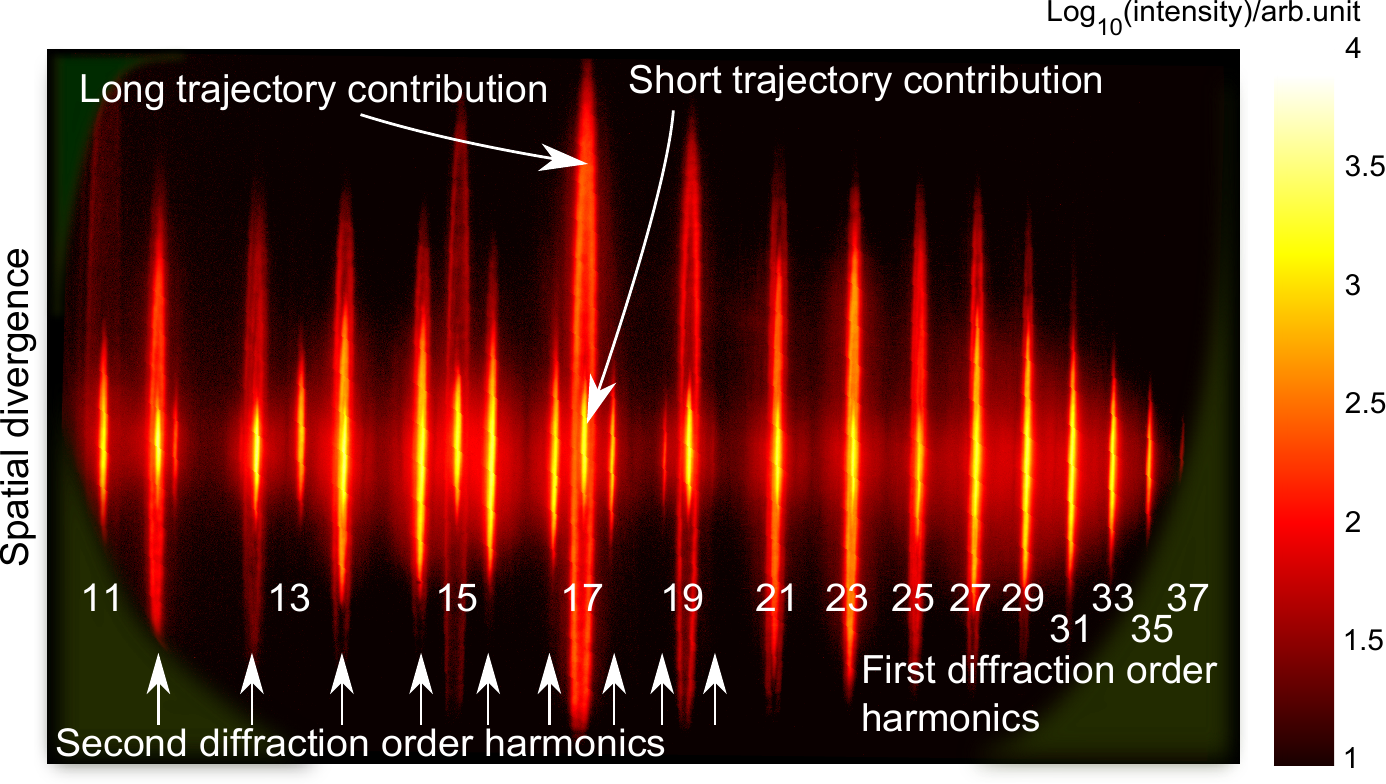}
  \caption{A typical high-order harmonic spectrum, generated from argon at 20\,kHz with 69.5\,$\upmu$J pulses. Examples of a short and a long trajectory electron contribution are indicated. The harmonics in the first diffraction order are identified and those in the second diffraction order indicated by arrows.}
  \label{fig: Argon image}
\end{figure}

Figure~\ref{fig: Argon image} shows a typical spectrum, obtained at 20\,kHz repetition rate in argon. Typical for harmonic spectra generated with 170\,fs long laser pulses is that the harmonics are spectrally well defined. This spectral resolution enables a very clear separation of two different contributions to each of the lower harmonics, coming from two different electron trajectories leading to the same final energy~\cite{LewensteinPRA1994}. Electrons escaping early after the maximum of the electric field experience a longer time in the continuum and follow a so-called long trajectory while those escaping later follow a shorter one. 

Since the long trajectory electrons spend more time in the continuum than the short trajectory electrons and therefore acquire more intensity-dependent phase, the generated light will have a more curved wavefront then the light from the short trajectory electrons. The emission is therefore more divergent than the emission from short trajectory electrons. This is clearly seen for harmonic orders 11--29 where the more divergent contribution from the long trajectory electrons surround a central spot being the contribution of the short trajectory electrons. For even higher harmonic orders (31 and higher), within the cut-off region, the two trajectories merge into one with a decreasing divergence as the energy increases.

\begingroup
\squeezetable
\begin{table}
\begin{ruledtabular}
\begin{tabular}{|p{1.5 cm}| p{1.5 cm}| p{1.5 cm} p{2 cm}|}
Repetition rate/kHz & Laser pulse energy/$\upmu$J & Harmonic order & Photon energy /eV\\
\hline
20 & $<$175 & 41 & 49.4\\
50 & $<$90& 35 & 42.1\\
100 & 54& 33 & 39.7\\ 
200 & 30 & 27 & 32.5\\ 
300 & 20 &  21 & 25.3\\
400 & 15& 19 & 22.9\\

\end{tabular}
\end{ruledtabular}
  \caption{The highest observed harmonic orders and corresponding photon energies generated in argon for different repetition rates and pulse energies.}
  \label{tab: HOHO for rep rate and pulse en}
\end{table}
\endgroup

Exploiting the variable repetition rate of the laser, we investigated the maximum repetition rate at which our setup can provide high-order harmonics in argon. Table~\ref{tab: HOHO for rep rate and pulse en} shows the highest observed harmonic order for various repetition rates and pulse energies. The generation conditions were individually optimized, with different iris opening diameters and with different positions of the gas jet.  As can be seen in the table, HHG up to a rate of 400\,kHz was possible.

\begingroup
\squeezetable
\begin{table*}
\begin{ruledtabular}
 \begin{tabular}{|l l || p{2.5cm} |p{2.5 cm} || p{2.5 cm} | p{2.5 cm}|}
 & & \multicolumn{2}{ c|| }{20\,kHz} & \multicolumn{2}{ c| }{100\,kHz}\\
 \hline

Harmonic order & Photon energy/eV & Number of photons generated/s $\cdot$ 10$^{11}$ & \mbox{Conversion efficiency} $\cdot$ 10$^{-7}$ & Number of photons generated/s $\cdot$ 10$^{10}$ & \mbox{Conversion efficiency} $\cdot$ 10$^{-8}$ \\
  \hline
13 & 15.6 & 0.5 & 0.3   &  -   &  -   \\
15 & 18.1 &  0.5 & 0.4  &  0.4 &  0.2 \\
17 & 20.5 &  2.3 & 2.2   &  3.0 &  1.9 \\
19 & 22.9 &  3.0 & 3.1  &  3.9 &  2.7 \\
21 & 25.3 &  4.2 &  4.9  &  4.1 &  3.1 \\
23 & 27.7 &  2.3 &  3.0 &  4.2 &  3.4 \\
25 & 30.1 &  2.3 &  3.1  &  4.4 &  3.9 \\
27 & 32.5 &  1.8 & 2.7   &  4.1 &  4.0 \\
29 & 34.9 &  1.4 & 2.2   &  3.5 &  3.7 \\
31 & 37.3 &  1.4 & 2.3  &  1.3 &  1.5 \\
33 & 39.7 &  0.9 & 1.7   &  0.1 &  0.1 \\
35 & 42.1 &  0.4 & 0.7   &  -   &  -   \\
37 & 44.5 &  0.1 & 0.2    &  -   &  -   \\
39 & 46.9 & 0.03 & 0.07    &  -   &  -   \\
\hline
Total & & 21.1& 27.0&  29.0& 24.3  \\

 \end{tabular}
\end{ruledtabular}
  \caption{The number of XUV photons per second as well as conversion efficiencies for the different harmonic orders generated in argon at 20\,kHz and 100\,kHz repetition rate. The pulse energy was 175\,$\upmu$J and 54\,$\upmu$J at the two repetition rates.}
  \label{tab: flux conv}
\end{table*}
\endgroup

In order to estimate the generated photon flux and conversion efficiency we have to correct for the losses induced in the measurement. Assuming a total aluminum oxide layer of 15\,nm \cite{LopezMartensPRL2005} on the aluminum filter \cite{HenkeADNDT1993} and compensating for the diffraction efficiency of the grating and the quantum efficiency of the XUV camera, the flux of the generated XUV photons in argon was estimated for 20 and 100\,kHz repetition rates. For each of these measurements the pulse picker was set to remove a number of laser pulses in order to not saturate the detection, and Table~\ref{tab: flux conv} lists the fluxes that would have been obtained, had all the laser pulses been used, together with the corresponding conversion efficiencies for each harmonic order. As can be seen in the table, up to 4.2$\cdot10^{11}$ photons/s per harmonic order were generated at 20\,kHz and a pulse energy of 175\,$\upmu$J, corresponding to a conversion efficiency of $4.9\cdot10^{-7}$. At 100\,kHz and a pulse energy of 54\,$\upmu$J, up to 4.4$\cdot10^{10}$ photons/s were generated per order, corresponding to a conversion efficiency $3.9\cdot10^{-8}$.

\begin{figure}
        \centering
        \begin{subfigure}[b]{0.4\textwidth}
                \includegraphics[width=\textwidth]{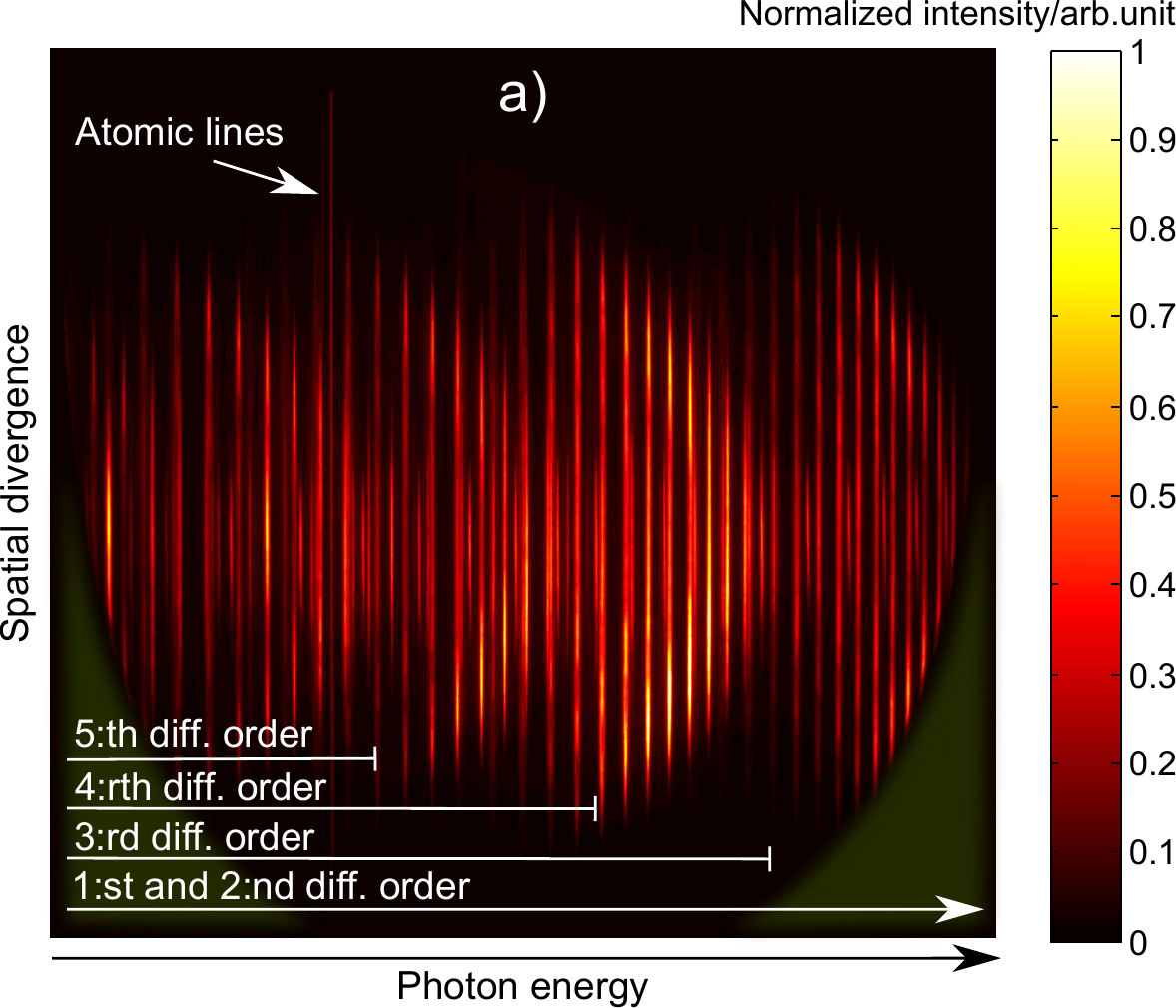}
                \label{fig: Neon2DSpectrum}
        \end{subfigure}%

        \begin{subfigure}[b]{0.4\textwidth}
                \includegraphics[width=\textwidth]{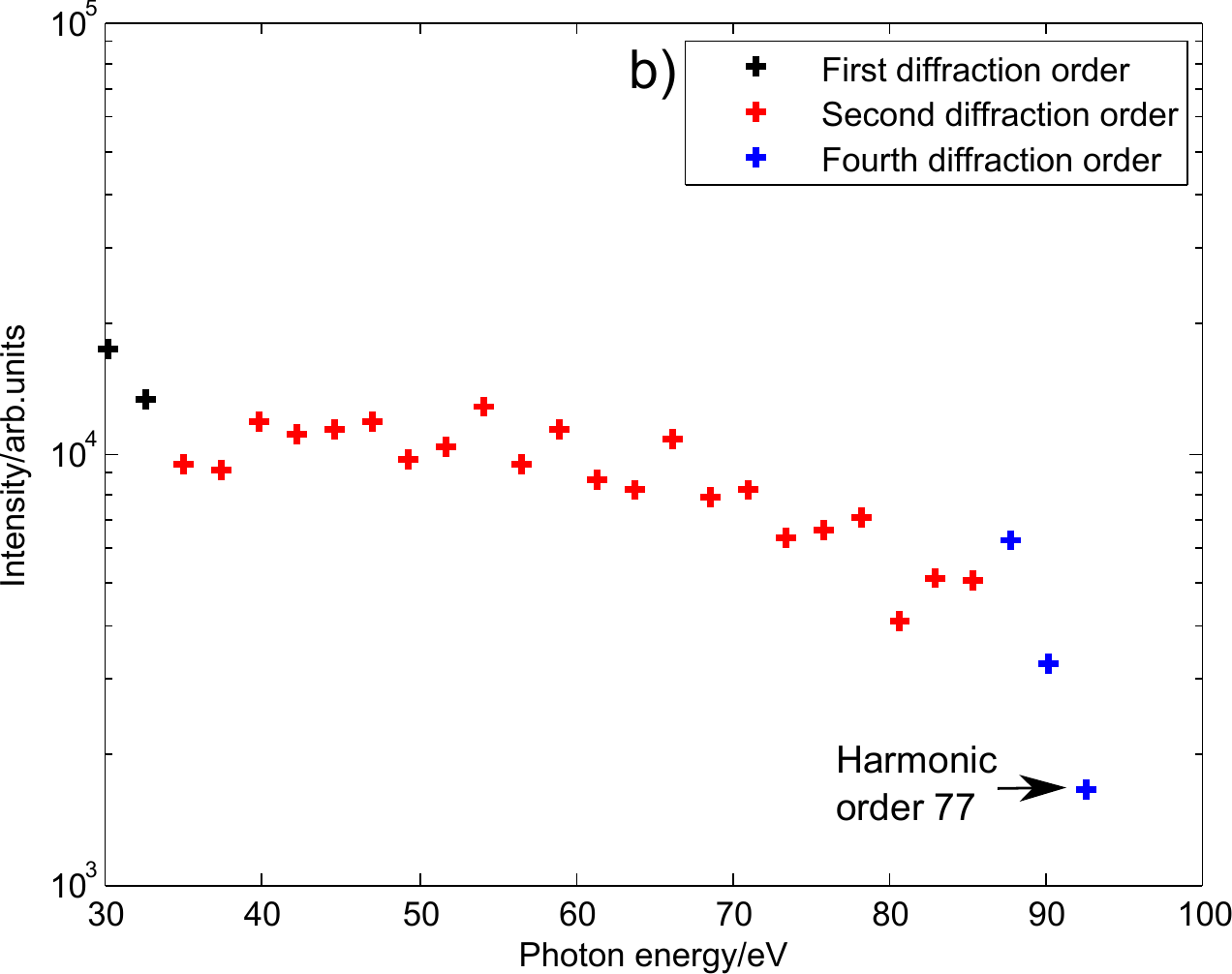}
                \label{fig:NeonBuiltSpectrum}
        \end{subfigure}
                \caption{A neon harmonic spectrum, obtained at 20\,kHz, is shown in a). The two atomic lines used for calibration are indicated, together with the diffraction orders present in the image. A lineout is performed on a) along the photon energy axis and at the center of the spectrum in the spatial divergence direction. The values of the peaks corresponding to the identified harmonic orders are shown in b). The efficiency of the grating is different for different diffraction orders and the harmonic amplitudes of the first and fourth diffraction orders were adjusted to fit the harmonic amplitudes of the second diffraction order. Photon energies above 90 eV can be observed.}
                     \label{fig:Neon Spectra}
\end{figure}

High-order harmonics were also generated in neon. Since neon has a higher ionization potential than argon (21.56\,eV compared to 15.76\,eV) the laser intensity can be increased without completely ionizing the gas and saturating the HHG process, and thereby higher cut-off energies are reached. The yield is, however, strongly reduced compared to when using argon.

Since our grating was not optimized for the short wavelengths obtained when generating harmonics in neon, several diffraction orders overlapped (see Fig.~\ref{fig:Neon Spectra}) and the harmonics had to be sorted into their respective diffraction order. The different diffraction orders can easily be identified from the change in divergence as the harmonic cut-off is approached (see Fig.~\ref{fig:Neon Spectra}). The cut-off part for both the first and the second diffraction order were outside the range of the detector and there is unfortunately a partial overlap between the harmonic orders in the first and the third diffraction orders. This is why we had to use the fourth diffraction order to identify the highest achievable harmonic order. The highest harmonic order we observed in neon was 77 (92.7\,eV or 13.4\,nm) obtained at 20\,kHz repetition rate with a pulse energy of 170\,$\upmu$J.

\section{Conclusion}
\label{Sec: Conclusion}
In conclusion we have demonstrated high-order harmonic generation using a high-repetition-rate turnkey laser. We have performed a technical investigation of our system and demonstrated that high-order harmonics could be generated at up to 400 kHz repetition rate in argon. A flux of up to 4.2$\cdot10^{11}$ and 4.4$\cdot10^{10}$ XUV photons/s per harmonic order was measured for 20\,kHz and 100\,kHz, respectively. This corresponds to the conversion efficiencies of $4.9\cdot10^{-7}$ and 3.9$\cdot10^{-8}$. High-order harmonic generation using this compact, user-friendly and stable laser was achieved in argon, neon and air. In neon harmonic orders as high as 77 were generated, which corresponds to 13.4\,nm or 92.7\,eV. This study shows that high-order harmonics, and hence attosecond pulses, can be generated at high repetition rate, flux and photon energy with a turnkey laser. This can be of interest to all attoscience experiments benefiting from high repetition rate HHG, but could also invite researchers from new fields to start using high-order harmonics.

\begin{acknowledgments}
This research was supported by the Swedish Foundation of Strategic research, the Marie Curie program ATTOFEL (ITN), the Swedish Research Counsil and the Knut and Alice Wallenberg Foundation.

\end{acknowledgments}

\hyphenation{Post-Script Sprin-ger}

\end{document}